\documentstyle[aps,twocolumn,epsf]{revtex}
\def\ave#1{\langle #1\rangle}

\newcommand{\braket}[2]{\langle #1|#2\rangle}

\newcommand{\sgn}{{\rm sgn}}
\begin{document} 
\title{Quantum localization and cantori in chaotic billiards}
\author{Giulio Casati}
\address{International Center for the Study of Dynamical Systems,
via Lucini,3, I--22100 Como, ITALY,\\ 
Istituto Nazionale di Fisica della Materia and INFN, Unit\`a di
Milano}
\author{Toma\v z Prosen}
\address{Physics Department, Faculty of Mathematics and Physics,
University of Ljubljana, Jadranska 19, 1111 Ljubljana, Slovenia
}
\date{\today}
\draft
\maketitle
\begin{abstract}
We study the quantum behaviour of the stadium billiard. We discuss how
the interplay between quantum localization  and the rich structure of the
classical phase space influences the quantum dynamics. The analysis of
this model leads to new insight in the understanding of quantum properties
of classically chaotic systems.
\end{abstract}

The study of quantum mechanics of complex systems at the light of the
chaotic behaviour of the corresponding classical systems has greatly
improved our understanding of quantum motion \cite{CC}. For example, the possibility put forward long
ago \cite{CGV} that
Random Matrix Theory may be a convenient tool to describe spectral
properties of classically chaotic systems rests now on more solid
grounds \cite{ALT,BOH}. However, in spite of the
progress in
recent years and the growing interest in the so-called ``quantum chaos'' 
we are still far from a satisfactory understanding, as a great variety and, to
some extent, unexpected rich behaviour of quantum motion continues to
emerge \cite{CMS} for which a satisfactory explanation is required. 
For example, the phenomenon of quantum dynamical localization  
discovered 20 years ago in systems under external
perturbations \cite{CCFI} and now experimentally confirmed 
\cite{Raizen}, mainly rests on numerical
computations and on qualitative considerations. Only few mathematical
results exist and it is not clear whether existing semiclassical
theories can account for this important feature.
Even less understood is the mechanism of dynamical localization in conservative systems\cite{CCGI93}.
Billiards are very convenient models to study since they can display a
rich variety of dynamical behavior from completely integrable (e.g. the
circle) to weakly mixing (e.g. the triangle with irrational angles)
\cite{ART}, to
completely chaotic with power law decay of correlations 
(the stadium), up to
exponential decay of correlations (dispersive billiards) \cite{ACG}. 
In addition they can studied numerically with great efficiency and
high accuracy (here we are able to compute accurate eigenvalues and 
eigenfunctions of the {\em desymmetrized} billiard with sequential 
number up to $10^7$).
They can be, to some extent, studied analytically
 and  in laboratory  experiments \cite{RIC,STO,SHR} .
Also they may be relevant for technological applications such as the
design of novel microlasers or other optical devices \cite{STONE}.

Recently localization has been shown to take place in the stadium
billiard\cite{BCL} and on other similar models
\cite{FS,CP,BORGO,CP2}. It is associated to the
fact that for small perturbations of the circle, the 
angular momentum undergoes a  classical diffusive process
and quantum effects may lead to suppression of this diffusive 
excitation.

In this paper we study the localization phenomenon and the structure 
of eigenfunctions (EF) as one moves from the perturbative regime to 
the ergodic, delocalized regime. The rich variety of classical phase
space determines a quite complicated quantum
structure. Indeed  the classical motion in the stadium billiard can 
be described by a discontinuous map of the  saw-tooth type. This map 
is known to have cantori \cite{DANA} which may act as barriers to
quantum motion \cite{GEISEL}. 
This effect has been discussed in \cite{MCKAY} and 
recently confirmed in numerical computations on the saw-tooth map
on the cylinder \cite{BORGO}.

\begin{figure}[htbp]
\hbox{\hspace{-0.1in}\vbox{
\hbox{
\leavevmode
\epsfxsize=3.5in
\epsfbox{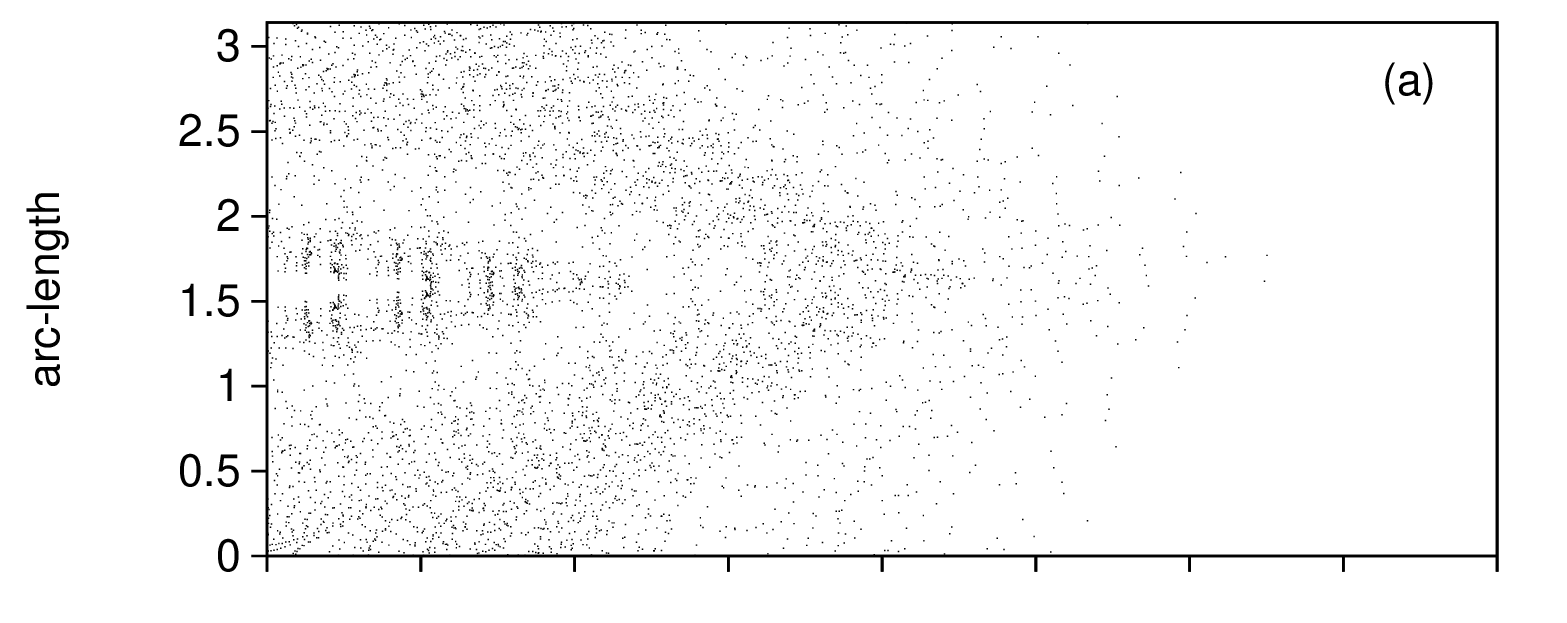}}
\vspace{-0.15in}
\hbox{
\leavevmode
\epsfxsize=3.5in
\epsfbox{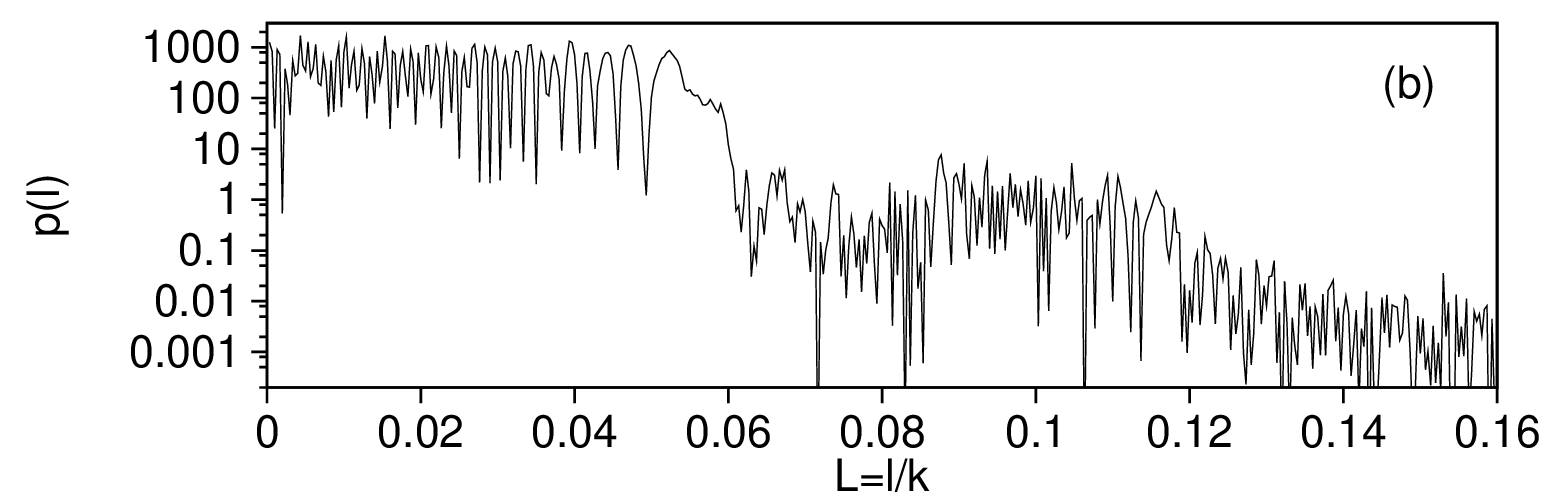}}
}}
\caption{
Time evolution of a single classical orbit, followed up to 
$20,000$ bounces, for the classical billiard with 
$\epsilon=0.003$. The orbit is initially started in the
middle of the largest `island' ( $L=0, \theta=\pi/4-0.0016$) 
(a). Angular momentum probability distribution $p_k(l)$
of the corresponding  eigenstate $\Psi_k$ with $\epsilon=0.003$ and 
eigenvalue $k=5999.8166$. As it is seen  the state is uniformly 
distributed over the cantorus in the main island (b).}
\label{fig:1}
\end{figure}

\begin{figure}[htbp]
\hbox{\hspace{-0.1in}\vbox{
\hbox{
\leavevmode
\epsfxsize=3.5in
\epsfbox{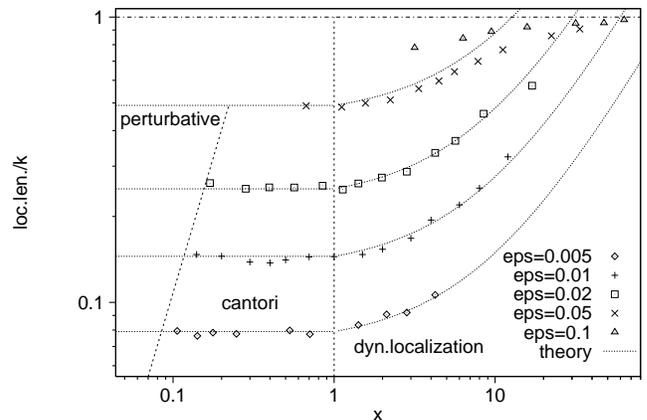}}
}}
\caption{ Rescaled localization length $\sigma$
versus the scaling variable $x=\epsilon^{3/2} k$ for five 
values of $\epsilon$ ($60 < k < 12,000$).
Each point is obtained by averaging over a large number $\nu$ of 
consecutive eigenstates ($\nu = 100$ for small $k$ and $\nu = 
1,000$ for large $k$). The numerical data clearly show the 
cantori border $x=1$. In the cantori region $\sigma$ is 
constant as expected, while for $x>1$
the numerical data agree with the theoretical prediction
(\ref{eq:res}) (dotted curves). For large $x$, the value of 
$\sigma$ approaches the maximal ergodic 
value $\sigma=1$.}

\label{fig:2}
\end{figure}

\begin{figure}[htbp]
\hbox{\hspace{-0.1in}\vbox{
\hbox{
\leavevmode
\epsfxsize=3.5in
\epsfbox{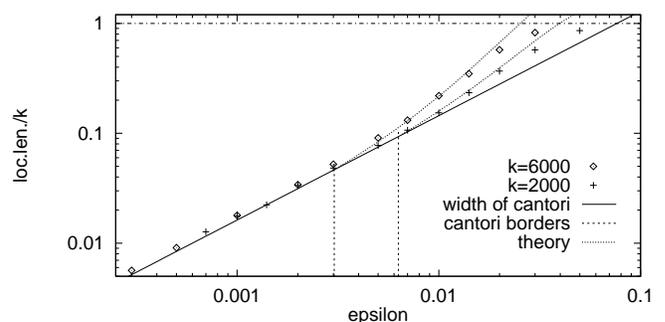}}
}}
\caption{
Rescaled localization length $\sigma$ versus 
$\epsilon$ for two different values of $k$.
The full lines give the classical estimate for the average 
width of  cantori. It is seen that below the cantori border $x=1$
($\epsilon=0.003$ for $k=6,000$ and $\epsilon=0.0063$ for 
$k=2,000$) $\sigma$ is proportional to $\epsilon$ and independent 
on $k$. Above the border $x=1$ instead, the numerical data follow 
the theoretical estimate (\ref{eq:res}) (dotted curves).}

\label{fig:3}
\end{figure}

In the following we discuss how the combined presence of the cantori
structure and of quantum
dynamical localization acts on EFs until the regime of quantum ergodic
behaviour is
reached.
We consider the motion of a free
point particle of unit mass and velocity $\vec{v}$ (energy
$E=v^2/2$) bouncing elastically inside a stadium-shaped well: 
two semicircles of radius $1$ connected by two straight line 
segments of length $2\epsilon$. 
The classical motion, for arbitrary small $\epsilon$, is ergodic, 
mixing and exponentially unstable
with Lyapunov exponent $\Lambda \sim \epsilon^{1/2}$.
It can be approximated (up to ${\cal O}(\epsilon)$)
by the discontinuous stadium-map \cite{BCL}
$
L_{n+1} = L_n - 2\epsilon \sin\theta_n\sgn(\cos\theta_n)\sqrt{1-L_n^2},
\theta_{n+1} = \theta_n + \pi - 2\arcsin{L_{n+1}}
$
for the rescaled angular momentum $L=l/\sqrt{2E}$ where $l =
\vec{r}\wedge\vec{v}$, and the polar angle $\theta$
(identical to arc-length for small $\epsilon$). From rigorous 
results on the sawtooth-map \cite{DANA} and from the stadium dynamics  
it can be shown that the angular momentum  (for 
small $\epsilon$) undergoes
a normal diffusive process with diffusion rate
\begin{equation}
D = \ave{(l_n - l_0)^2}/n\vert_{n\gg 1}
\approx 2 \epsilon^{5/2} (2E - l^2)
\label{eq:diff}
\end{equation}
(Notice that diffusion rate $D$  depends on the local value of angular momentum).
The power $5/2$ in eq. (\ref{eq:diff}) 
is due to the existence of {\em cantori}
which form strong obstacles to phase space transport.
In ref \cite{BCL} the phenomenon of quantum localization has been 
shown to take place in the stadium billiard leading to strong deviations from RMT predictions.
However the dependence of the localization length on system parameters
is not known and in particular we do not know if, and to what extent, the
presence of cantori will influence the quantum dynamics. Indeed 
it has been conjectured \cite{MCKAY}
that  cantori act as perfect barriers for quantum motion provided
the flux through cantori is smaller than a Planck's cell, ${\cal F} < 2\pi\hbar$.
On the basis of results on the sawtooth-map \cite{DANA} we can estimate the 
{\em flux} ${\cal F}$ ---
phase space area transported through cantori per iteration
(bounce with the boundary) --- which is here independent of the winding number 
of the resonance, and, for small $\epsilon$, it is given by
${\cal F} \approx (2 E)^{1/2}\epsilon^{3/2}$
which leads to the  {\em cantori border} 
\begin{equation}
\epsilon_c = k^{-2/3},
\end{equation}
where $k=\sqrt{2E}$ is the wavenumber. 
For $ \epsilon < \epsilon_c$, ($x := k \epsilon^{3/2} < 1$), 
cantori act as perfect barriers, and the quantum system looks as 
if classically integrable. It is therefore expected that the 
localization length $\ell$ of eigenstates in angular momentum 
variable $l$ must be of the order of the size of cantori. 
This size, in rescaled angular momentum variables, 
averaged over all the resonances, can be estimated from the
exact results on sawtooth map \cite{sizeofcantori}, namely
$\bar{p}= c\epsilon$, where $c\approx 12$ is a numerical constant. 
($c=10$ for $\epsilon=0.05$, and 
$c=15$ for $\epsilon=0.005$). The fact that $c$ slowly 
increases with decreasing $\epsilon$ is due to the presence of 
the cantorus along the separatrix of 2:1 resonance 
(around $L=0$) which has a larger size, $p(2,1)\approx\sqrt{\epsilon}$.
In Fig.\ref{fig:1} we show the classical structure of cantori 
($\epsilon=0.003$) in phase space around the largest 2:1 resonance 
and the associated quantum eigenstate.  
In this regime the (average) rescaled localization length of
eigenstates $\sigma = \ell/\ell_{\max} = \ell/k$ 
is indeed found to be equal to the
(average) size of cantori (see Fig.\ref{fig:2},Fig.\ref{fig:3}),
$\sigma = \bar{p}$.

For $\epsilon > \epsilon_c$, ($x = k \epsilon^{3/2} > 1$),
when the flux trough turnstiles becomes larger than one quantum $2\pi\hbar$, 
the cantori do not act any more as barriers for quantum dynamics 
and the quantum motion starts to follow the classical diffusive 
behaviour up to the quantum relaxation time $t_R$ (break time), 
which is proportional to the density of 
{\em operative eigenstates}. For $t>t_R$ instead, the quantum dynamics 
enters an oscillatory regime around the stationary localized state with 
a localization length $\ell$. Therefore,
 $t_R = \sigma dN/dE$, where $\sigma=\ell/\ell_{\rm max}=\ell/k$,
and  $N=E/8$ (from Weyl formula). From the diffusion law (1) we have
 $\ell^2  \approx D t_R/T=\sigma D/T$, 
where $T\approx E^{-1/2}$ is the average time between  
bounces. Thus we obtain a simple expression for the rescaled 
averaged localization length $\sigma$
\begin{equation}
\sigma \approx D/k \approx \alpha \epsilon^{5/2} k = \alpha
\epsilon x.
\label{eq:dynloc}
\end{equation}
where $\alpha\approx 1.7$ is a numerical constant. However, 
to the above expression, we need to add the average size $\bar{p}$
of cantori. Therefore, for $x>1$, the actual expression for the localization length  will be given by
\begin{equation}
\sigma = \bar{p} + (1-\bar{p})\alpha \epsilon(x-1)
\label{eq:res}
\end{equation}
which takes into account also the fact that we need to rescale the
total size of angular momentum space,  and that for 
$x= 1$, $\sigma = \bar{p}$.
Eigenstates become delocalized  (ergodic) when 
$\sigma=1$, giving the ergodicity border $\epsilon = 
\epsilon_e$,
\begin{equation}
\epsilon_e \approx (\alpha k)^{-2/5}
\end{equation}
which agrees with the results of \cite{BCL}.
The {\em cantori border} can actually be observed
if it is below  the ergodic border and above the perturbative border
$\epsilon_p$. The {\em perturbative border} $\epsilon_p$ is 
given by the condition that $\epsilon$ should be large enough,
$\epsilon > \epsilon_p$, to couple two neighboring eigenvalues of
angular momentum, which is equivalent to the intuitive condition
of comparing the deformation $\epsilon$ and the De Broglie wavelength,
which leads to
\begin{equation}
\epsilon_p = k^{-1}.
\end{equation}
Therefore, for sufficiently large $k$, we have, 
$\epsilon_p < \epsilon_c < \epsilon_e$. In this situation it is natural 
to expect that cantori will influence the localization process and 
we may have here a nice possibility to study the effect of cantori
in quantum mechanics.

In order to check the above predictions we numerically 
computed quantum eigenfunctions $\Psi_k(\vec{r})$ of the
stadium billiard (solutions of the Schr\" odinger equation
$(\nabla^2 + k^2)\Psi_k = 0$ where $\hbar=1$) by expanding them 
in terms of circular waves (here we consider only odd-odd states)
$\Psi_k(\vec{r}) = \sum_{s=1}^{M} a_s J_{2s}(kr)\sin(2s\theta)$.
Eigenvalues $k=k_n$ and the associated coefficients $a_s$ have been 
computed very efficiently \cite{CP2} by minimizing a special quadratic 
form defined along the boundary of the billiard \cite{VS95}. 
The coefficient $a_s$ is proportional to the probability amplitude 
of finding angular momentum equal to $l=2s$, 
$p_k(l)=|\braket{l=2s}{\Psi_k}|^2 = |a_s|^2 \int_0^1 dr r 
|J_{2s}(kr)|^2 \propto |a_s|^2\sqrt{1-l^2/k^2}$.

Quantum localization in the stadium is not exponential like for kicked 
rotor or smooth diffusive billiards \cite{FS}, but rather of 
power-law type. More precisely, the tails of eigenfuctions have 
been found\cite{CP2} to decay, on average, as 
$p(l) \sim |l-\ave{|l|}|^{-4}$. 
Therefore,  in order to characterize the localization length of quantum eigenfunctions $\Psi_k$,
we choose the $99\%$ probability localization length 
$\sigma$ rather than the more common inverse participation ratio, information
entropy or uncertainty of angular momentum; indeed, the former is 
quite independent of the nature of tails of the distribution $p(l)$ 
unlike the latter quantities. More precisely, we define 
$\sigma_k$ as the minimal number of 
angular momentum states which are needed to support the  
$99\%$ probability of an eigenstate $\Psi_k$, 
$\sigma_k = \min\{\#{\cal A},\sum_{l\in{\cal A}} p_k(l) 
\ge 0.99\}/(0.725 k)$.  The normalization factor has been chosen 
in such a way that $\sigma_k=1$ for completely delocalized (GOE) states.
In Fig.2 and Fig.3 we show the 
dependence of the averaged rescaled localization length 
$\sigma=\ave{\sigma_k}$ (averaged over a set of consecutive 
eigenstates $\Psi_k$ with $k$ on a narrow interval) on the parameter 
$\epsilon$ and on the wavenumber $k$ 
(up to $k=12000$, $N\approx 10^7$).
Numerical data agree with theoretical predictions (\ref{eq:res}) 
with $\alpha=1.7$.

The exact stadium-eigenstates 
$\Psi_{k_n}$ may be expanded in terms of unperturbed 
quarter-cirlce eigenstates $\Phi_{s m}(\vec{r}) =
J_{2s}(k^0_{sm}r)\sin(2s\theta)$ where $k^0_{sm}$ are the
eigenvalues of the integrable quarter-circle -- the zeros of the
even-order Bessel functions.
The matrix $c^{n}_{sm} = \braket{\Phi_{sm}}{\Psi_{k_n}}$ 
can be easily computed from the coefficients $a_l$ \cite{FS}.

It is important to note that unlike for Wigner band random 
matrices \cite{CCGI93} the matrix $c^{n}_{sm}$, 
(ordered, as usual, with increasing quantum wavenumbers 
$k_n, k^0_{sm}$), has a symmetric appearance 
\cite{CP2}. It has a band structure with the bandwidth $b\sim k=\delta E$, 
independent of $\epsilon$ for $\epsilon < \epsilon_p$, 
where $\delta E $ is the width of the energy shell.
However, below the ergodicity border, the
matrix $c^n_{sm}$ is uniformly {\em sparse}, both in
horizontal and vertical directions. The effective
number of nonzero elements in each row (or column) is
(on average) equal to the localization length $\ell = \sigma 
k$. In order to illustrate the general structure we show in Fig.4
the probability distributions of a typical circle state in terms of 
eigenstates of the stadium (one column of the matrix $|c^n_{sm}|^2$)
and a typical eigenstate of the  stadium in terms of circle
states (one row of the matrix). These distributions are
strongly sparse and extended over the whole energy shell of $\delta E
\sim b=k$ states. In Fig.4c we show, for the same stadium 
eigenstate, the probability 
distribution $p(l)$ in angular momentum: as expected it is
 strongly localized and 
non-sparse. The above structure is found in the
regime of dynamical localization and in the regime of cantori 
localization where the quantum system behaves as if classically 
integrable\cite{PRA}. Furthermore, a similar structure is found also for the 
nearly-circular but chaotic rough billiards with smooth 
boundaries introduced in \cite{FS} (above the
`Breit-Wigner regime'), where the situation is even more complicated due 
to KAM structure of classical phase space. 

As the parameter $\epsilon$ is increased up to the ergodicity border
$\epsilon_e$, sparsity decreases and the quantum
angular momentum distribution $p(l)$ approaches (apart from fluctuations) 
the classical steady-state microcanonical distribution, 
$p_e(l) \propto \sqrt{2E - l^2}$.
Notice that the scaling parameter $\sigma$ also controls the deviations
from RMT predictions of the statistical properties of eigenvalues and 
eigenfunctions.

In this paper we have discussed a dynamical model, the
stadium billiard, for which the classical motion is completely
chaotic (without any island of stability) for any value of the control 
parameter $\epsilon$ and we have shown that the quantum dynamics instead,
exhibits a rich structure and different
regimes of motion as a function of $\epsilon$ and energy $E$.
It has been shown that the presence of {\em cantori} in classical phase
space may have strong effect on the quantum dynamics and
leads to a new border which is different from 
the perturbative and the ergodic border.
In the regime of quantum cantori (where the phase space
flux through cantori is less than one quantum)
the rescaled localization length $\sigma=\ell/k$ does not depend
on energy or wavenumber $k=\sqrt{2E}$.
However, above the cantori border, quantum dynamical
localization takes place and  the localization length $\ell$ is found to be 
proportional to the rate $D$ of classical diffusion in angular 
momentum. The mechanism of localization is strongly connected
to the sparsity of  EFs when expanded on the basis of
(unperturbed) circle states (and  vice versa). 
We suggest that the above features are typical of the quantum dynamics of 
classically chaotic conservative systems.

Discussions with R.Prange, R.S.MacKay, J.Keating, G.Tanner and I.Dana
are gratefully acknowledged. T.P. acknowledges financial support from 
the Ministry of Science and Technology of R Slovenia.

\begin{figure}[htbp]
\hbox{\hspace{-0.1in}\vbox{
\hbox{
\leavevmode
\epsfxsize=3.5in
\epsfbox{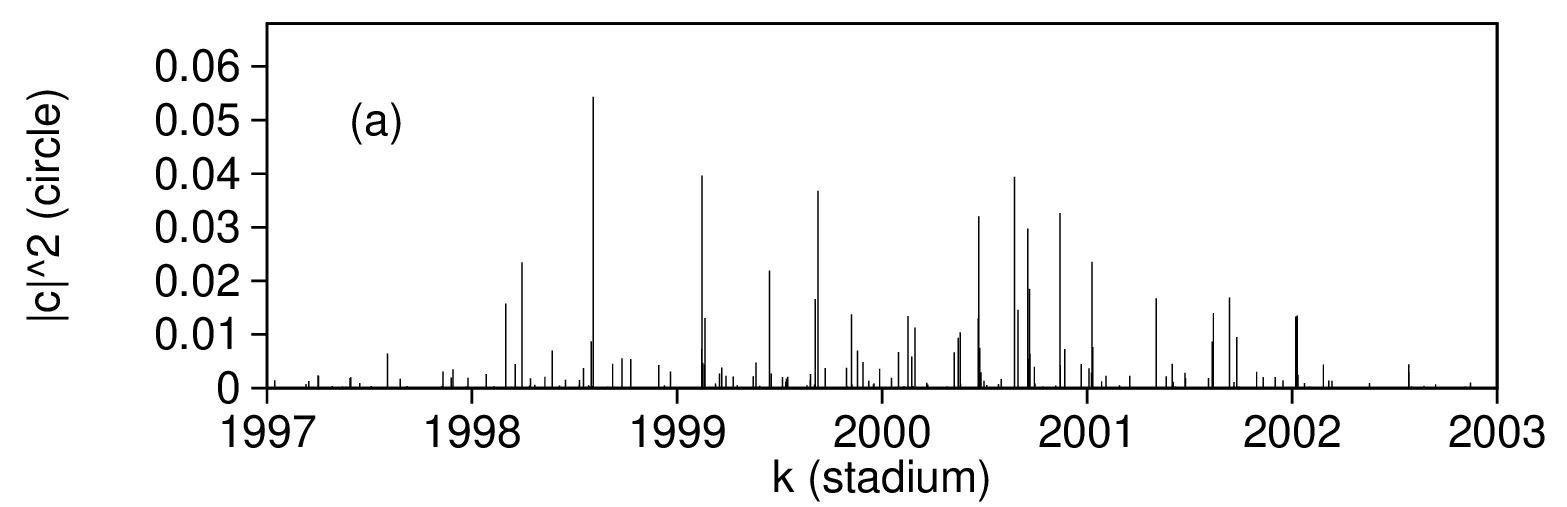}}
\hbox{
\leavevmode
\epsfxsize=3.5in
\epsfbox{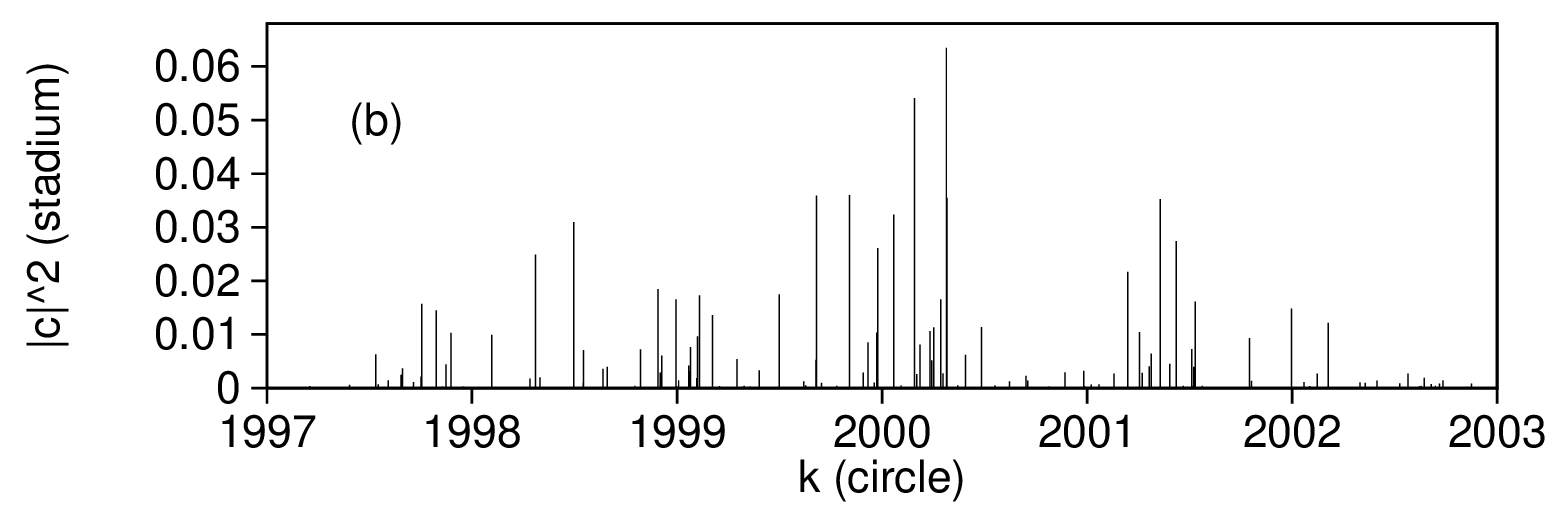}}
\hbox{
\leavevmode
\epsfxsize=3.5in
\epsfbox{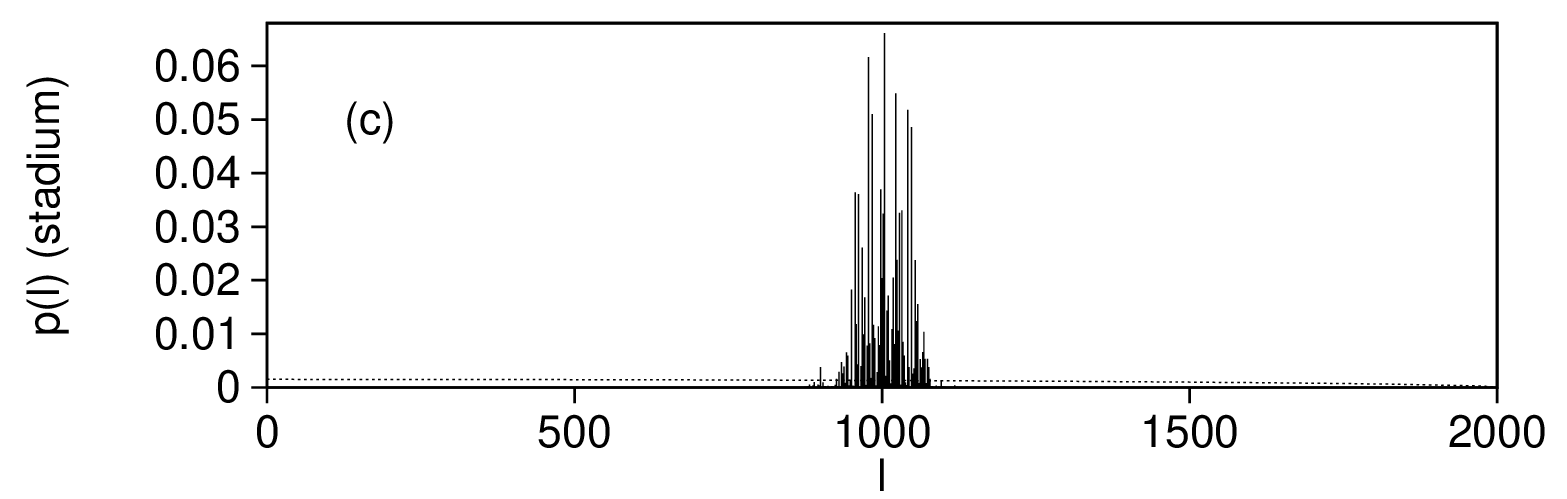}}
}}

\caption{
The structure of high-lying eigenstates of the stadium 
$k_n\approx 2000$ for $\epsilon=0.01$ in terms of circle states 
and vice versa. In (a) we show the probability distribution of a 
typical circle state $(l=2s=848,k^0_{sm}=1999.99349)$ 
versus the wavenumber (energy) of the stadium eigenstates  
(a column of the matrix $|c^{n}_{sm}|^2$ with 
fixed $s,m$), in (b) we show the probability distribution of a 
typical eigenstate of the stadium ($k_n=1999.91397$) versus the 
wavenumber of circle states (a row of the matrix $|c^{n}_{sm}|^2$
with fixed $n$).  In (c) we show the same state 
as in (b) but in angular momentum quantum number $l$ (the sum 
over radial quantum number $\sum_m |c^{n}_{sm}|^2$) .
Notice that the rescaled localization length here is $\sigma_{k_n}=0.10$
and, correspondingly, only a fraction $\sigma_{k_n}$ of the total number of
$\sim k$ states inside the energy shell is actually significantly
excited (a,b). The dotted curve in (c) gives the ergodic 
distribution $p_e(l)=(8\pi/k^2)\sqrt{k^2 - l^2}$.
}
\label{fig:4}
\end{figure}


\vspace{-1cm}

\begin{thebibliography}{99}
\bibitem {CC} G.Casati and B.V.Chirikov, eds., {\em Quantum Chaos:
Between Order and Disorder}, Cambridge University Press (1994);
Physica {\bf D86}, 220 (1995).

\bibitem{CGV} G. Casati, I Guarneri and F. Valz-Gris, Lett. Nuovo
Cimento{\bf 28},279 (1980); E. McDonald and A.N. Kaufman, Phys.Rev.
Lett.{\bf42}, 1189 (1979).

\bibitem{ALT} A.V.Andreev, O.Agam, B.D.Simons, and B.L.Altshuler,
Phys.Rev.Lett.{\bf 76}, 3947 (1996).

\bibitem{BOH} O.Bohigas in Proceedings of the 1989 Les Houches
Smmer School on ``Chaos and Quantum Physics'', ed. by
M.J.Giannoni, A.Voros, J.Zinn-Justin (Elsevier Science Publisher
B.V., North -Holland, Amsterdam 1991), p 89.

\bibitem{CMS}G. Casati, G. Maspero and D.L. Shepelyansky: quantum strange
attractor preprint.

\bibitem{CCFI} G.Casati, B.V.Chirikov, J.Ford and
F.M Izrailev, Lecture Notes in Physics {\bf 93},
334 (1979).

\bibitem{Raizen} F.L. Moore, J.C. Robinson, C. F.Bharucha, B. Sundaram,
and M.G. Raizen Phys. Rev. Lett. 75, 4598(1995).

\bibitem{CCGI93} G. Casati, B.V. Chirikov, I. Guarneri and F.M. Izrailev,
Phys. Rev. E {\bf 48}, R1613 (1993); Physics Letters A223 (1996) 430.

\bibitem{ART} R. Artuso, G. Casati, and I. Guarneri, Phys. Rev E
{\bf55}, 6384, (1997). 

\bibitem{ACG} R. Artuso , G. Casati and I. Guarneri, J. Stat. Phys.
{\bf83}, 145, (1996).

\bibitem{RIC} H. Alt, H.-D. Graf, R. Hofferbert, C. Rangacharyulu, H. Rehfeld, 
A. Richter, P. Schardt, and A. Wirzba, Phys. Rev. E {\bf54}, 2303, (1996).

\bibitem{STO} J. Stein and H.-J. Stockmann, Phys. Rev. Lett.
{\bf68}, 2867, (1992).

\bibitem{SHR} A. Kudrolli, V. Kidambi, and S. Sridhar, Phys. Rev. Lett.
{\bf75}, 822 (1995).

\bibitem{STONE} J. U. Nockel and A.D.Stone, Nature (London) {\bf385},
45 (1997).

\bibitem{BCL}  F.Borgonovi, G.Casati and B.Li, Phys. Rev. Lett.
{\bf 77}, 4744 (1996).

\bibitem{FS} K.Frahm and D.Shepelyansky, Phys.Rev.Lett. {\bf 78},
1440 (1997); ibid. {\bf 79}, 1833 (1997).

\bibitem{CP} G.Casati and T.Prosen, preprint, cond-mat/9704084.

\bibitem{BORGO} F. Borgonovi, preprint, chao-dyn/9801032

\bibitem{CP2} G.Casati and T.Prosen, preprint, submit.to Physica D

\bibitem{DANA} Q. Chen, I. Dana, J. D. Meiss, N. W. Murray, and I. C. Percival,
Physica {\bf D 46}, 217 (1990);
see also I. Dana, N. W. Murray, and I. C. Percival,
Phys.Rev.Lett {\bf 62}, 233 (1989).

\bibitem{GEISEL} T. Geisel, G. Radons, and J. Rubner,
Phys.Rev.Lett {\bf 57}, 2883 (1986).

\bibitem{MCKAY} R.S. MacKay and J.D. Meiss, Phys. Rev. A37 (1988) 4702.

\bibitem{sizeofcantori}
Average the expression for $h(m,n)$, pp225,
weigted by phase-space areas of resonances (eq.33) of \cite{DANA}.

\bibitem{VS95} E.Vergini and M.Saraceno, Phys.Rev.E {\bf 52}, 2204 (1995).

\bibitem{PRA} This integrable-like behaviour maybe at the root of the 
analytical results in: R.E.Prange and R.Naverich, 
``Quasiclassical surface of section perturbation theory'', preprint.

\end{thebibliography}
\end{document}